# Luminosity Upgrades for ILC

Mike Harrison (Brookhaven), Marc Ross (SLAC), Nicholas Walker (DESY)

August 15, 2013

## Abstract

The possibility of increasing the luminosity for the ILC at $E_{cm}$ ≤ 350 GeV centre-of-mass by increasing the beam power are considered. It will be shown that an approximately constant luminosity can be achieved across the centre of mass energy range (250–500 GeV) without exceeding the installed AC power for 500 GeV operation. Overall a factor of four in luminosity over the published baseline could be achieved at 250 GeV resulting in $3 \times 10^{34}$ cm$^{-2}$s$^{-1}$. The implications for the damping rings and positron source are also briefly discussed.

## 1. Introduction

The ILC TDR [1] outlines two upgrade scenarios:

- a *luminosity upgrade* which doubles the average beam power by adding RF to the linacs;
- an *energy upgrade* to double the centre of mass energy to 1 TeV by extending the main linacs.

The TDR also describes a possible first stage 250 GeV centre of mass energy 'Light Higgs Factory', which assumes only 50% of the baseline linacs length are installed in an initial operations phase. These options are summarised in Table 1 which is taken directly from the TDR [1].

The proposed luminosity upgrade doubles the beam power by increasing the number of bunches per pulse by a factor of two from 1312 to 2625, resulting in a direct doubling of the baseline luminosity to $3.6 \times 10^{34}$ cm$^{-2}$s$^{-1}$ at a centre of mass energy of 500 GeV. The upgrade requires two major modifications to the baseline machine:

- Approximately 50% more klystrons and modulators must be installed in the main linacs.
- Depending on the limits imposed by the electron cloud instability, a second positron damping ring may need to be installed to accommodate the increased number of bunches. (The single existing electron damping ring should be able to accommodate the factor of two in beam current.)

Both the main linacs and the damping rings are already configured in the baseline design to facilitate the upgrades.  Additional klystrons and modulators are also required in the sources to accelerate the higher beam current. The electron ring RF systems will also require upgrading to provide the power for the higher currents. The upgrade remains relatively straightforward and within the scope of all the machine hardware specifications, and no significant R&D is required. As Table 1 indicates, the upgrade is estimated to increase the AC wall plug power by approximate 25% to 204 MW (at 500 GeV centre of mass).

Table 1: Upgrade and staging options for the ILC (beyond the baseline 500 GeV centre of mass machine). Taken directly from the TDR [1].

| | | | Baseline | 1st Stage | L Upgrade | TeV Upgrade A | TeV Upgrade B |
|---|---|---|---|---|---|---|---|
| Centre-of-mass energy | $E_{CM}$ | GeV | 500 | 250 | 500 | 1000 | 1000 |
| Collision rate | $f_{rep}$ | Hz | 5 | 5 | 5 | 4 | 4 |
| Electron linac rate | $f_{linac}$ | Hz | 5 | 10 | 5 | 4 | 4 |
| Number of bunches | $n_b$ | | 1312 | 1312 | 2625 | 2450 | 2450 |
| Bunch population | $N$ | $\times 10^{10}$ | 2.0 | 2.0 | 2.0 | 1.74 | 1.74 |
| Bunch separation | $\Delta t_b$ | ns | 554 | 554 | 366 | 366 | 366 |
| Pulse current | $I_{beam}$ | mA | 5.79 | 5.8 | 8.75 | 7.6 | 7.6 |
| Average total beam power | $P_{beam}$ | MW | 10.5 | 5.9 | 21.0 | 27.2 | 27.2 |
| Estimated AC power | $P_{AC}$ | MW | 163 | 129 | 204 | 300 | 300 |
| RMS bunch length | $\sigma_z$ | mm | 0.3 | 0.3 | 0.3 | 0.250 | 0.225 |
| Electron RMS energy spread | $\Delta p/p$ | % | 0.124 | 0.190 | 0.124 | 0.083 | 0.085 |
| Positron RMS energy spread | $\Delta p/p$ | % | 0.070 | 0.152 | 0.070 | 0.043 | 0.047 |
| Electron polarisation | $P_-$ | % | 80 | 80 | 80 | 80 | 80 |
| Positron polarisation | $P_+$ | % | 30 | 30 | 30 | 20 | 20 |
| Horizontal emittance | $\gamma\epsilon_x$ | µm | 10 | 10 | 10 | 10 | 10 |
| Vertical emittance | $\gamma\epsilon_y$ | nm | 35 | 35 | 35 | 30 | 30 |
| IP horizontal beta function | $\beta_x^*$ | mm | 11.0 | 13.0 | 11.0 | 22.6 | 11.0 |
| IP vertical beta function (no TF) | $\beta_y^*$ | mm | 0.48 | 0.41 | 0.48 | 0.25 | 0.23 |
| IP RMS horizontal beam size | $\sigma_x^*$ | nm | 474 | 729 | 474 | 481 | 335 |
| IP RMS veritcal beam size (no TF) | $\sigma_y^*$ | nm | 5.9 | 7.7 | 5.9 | 2.8 | 2.7 |
| Luminosity (inc. waist shift) | $L$ | $\times 10^{34}$ cm$^{-2}$s$^{-1}$ | 1.8 | 0.75 | 3.6 | 3.6 | 4.9 |
| Fraction of luminosity in top 1% | $L_{0.01}/L$ | | 58.3% | 87.1% | 58.3% | 59.2% | 44.5% |
| Average energy loss | $\delta_{BS}$ | | 4.5% | 0.97% | 4.5% | 5.6% | 10.5% |
| Number of pairs per bunch crossing | $N_{pairs}$ | $\times 10^3$ | 139.0 | 62.4 | 139.0 | 200.5 | 382.6 |
| Total pair energy per bunch crossing | $E_{pairs}$ | TeV | 344.1 | 46.5 | 344.1 | 1338.0 | 3441.0 |

While the TDR only explicitly discusses this luminosity upgrade scenario at 500 GeV centre of mass operation, it is in principle directly applicable to lower centre of mass energies above 300 GeV. Below 300 GeV some additional modifications to the damping rings are necessary to accommodate the shorter damping time required by the so-called 10-Hz mode operation, where the electron linac is pulsed at 10 Hz with alternate pulses being used to generate positrons with a 150 GeV beam. Operation of the single electron damping ring with high current and 100 ms damping time was not considered in the TDR, but as we will see this can be accommodated within the existing design with some risk (see Section 3).

Beyond the luminosity scenario outlined in the TDR, it is possible to consider a further increase in beam power (and therefore luminosity) by increasing the pulse repetition rate of the machine, from 5 Hz up to a maximum of 10 Hz (the current design specification of the main linac components). This is particularly attractive at lower centre-of-mass energies where the main linacs are used at reduced accelerating gradient, at which point the repetition rate can be increased to the limit of the installed AC power and cooling capacities (both conventional and cryogenic cooling). Running at 250 GeV centre-of-mass (light Higgs) with the linacs running at approximately half their nominal (maximum) gradient reduces their power consumption by over a factor of two, immediately opening up the potential of running the accelerator at 10 Hz collisions, doubling the luminosity. Combining this with TDR luminosity upgrade could then result in an overall factor of four improvement (3×10$^{34}$ cm$^{-2}$s$^{-1}$). Operation with 10 Hz collisions at 250 GeV centre of mass does however exclude the possibility of the 10 Hz mode positron production, and alternatives would have to be considered. The issue of positron production is discussed in Section 4.

The remainder of this report discusses in more detail the possibility of these luminosity enhancements. Section 2 discusses the main linac operation. Sections 3 and 4 discuss the implications and possible risks for the damping rings and positron source respectively. Finally Section 5 summarises the possible luminosity performance and required AC power, based on the implementation of the full 500 GeV machine, and Section 6 briefly discusses the implications of high current, high repetition rate running for a first-stage 250 GeV centre of mass machine (light Higgs factory).

## 2. Linac operation at lower gradient and higher repetition rates

For $E_{CM} \leq 500$ GeV operation, we assume that the beam energy is reduced by globally reducing the accelerating gradient in the main linacs down from the nominal average of *G* = 31.5 MV/m. The reduction in RF power is further aided by an increase in the RF-to-beam-power efficiency, due to the shorter cavity fill time (proportional to gradient). With the high beam current of 8.8 mA (TDR luminosity upgrade), the efficiency increases from ~60% to ~77% for 31.5 MV/m and 15.3 MV/m respectively. The other important constraint in the cryogenic cooling capacity, the dynamic part of which scales roughly as $G^2$ assuming a constant quality factor[1] ($Q_0$). Figure 1 shows the relative AC power for the RF and cryogenic systems as a function of centre of mass energy (linac gradient). The curves are scaled from the TDR 500 GeV power consumptions using simple scaling laws [2]. We can see that the repetition rate can be safely increased from the nominal 5 Hz to 8 Hz at the top threshold (350 GeV CM) and 10 Hz at the light Higgs (250 GeV CM), without exceeding the installed power[2].

Figure 2 shows how the total power for the main linac scales with centre of mass energy (gradient). Again the possible increase in repetition rate is indicated for 350 GeV and 250 GeV centre-of-mass operation. It should be noted that the 130 MW indicated for the 500 GeV, 5 Hz operation, from which the all other values are scaled, only represents that part which changes with gradient. The total main linacs power requirement is slightly higher than this (~142 MW).

---

[1] This is likely to be a conservative assumption, as results indicate factors of 2 to 4 higher $Q_0$ values at lower gradients.
[2] In principle the repetition rate could be increase to 14% Hz at E$_{cm}$ = 250 GeV.

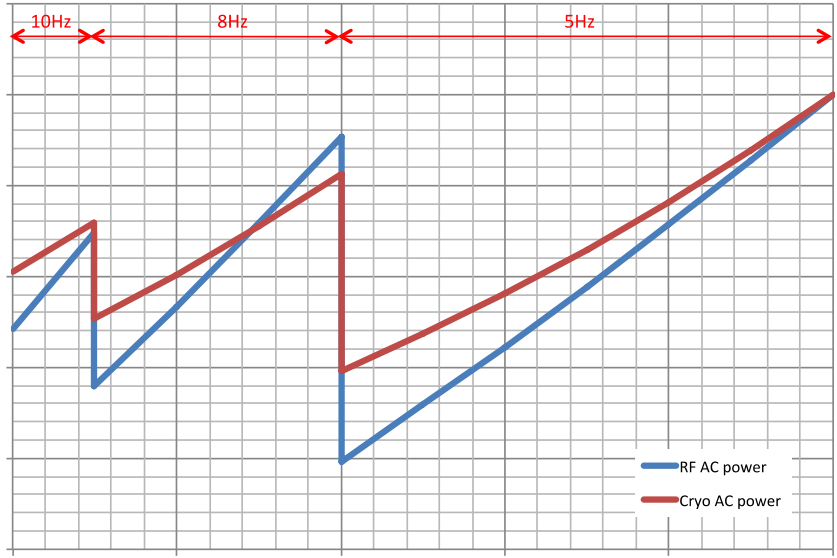

**Figure 1:** AC power associated with the main linacs' RF and cryogenic systems, relative to the installed values at 500 GeV centre of mass operation. An increase in pulse repetition rate at 350 GeV (8 Hz) and 250 GeV (10 Hz) centre-of-mass is indicated.

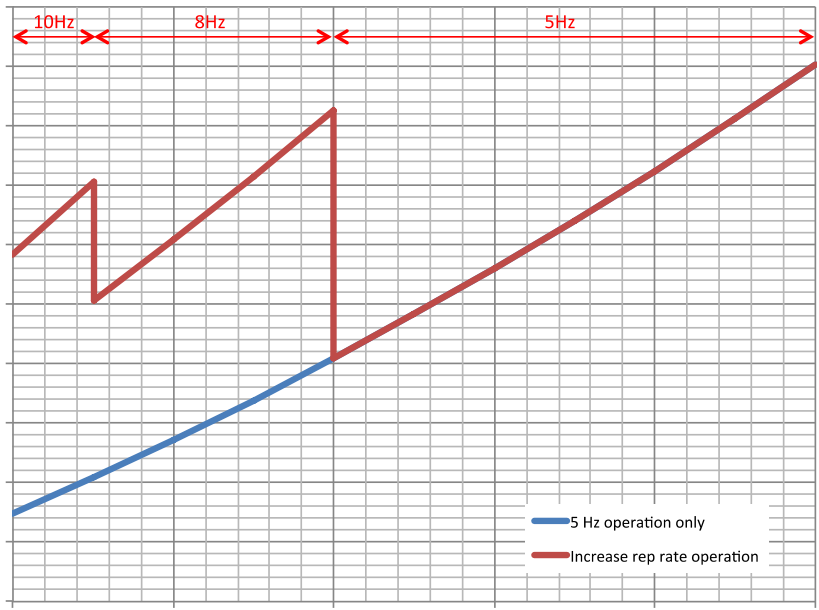

**Figure 2:** AC power for the main linacs as a function of centre-of-mass energy operation.

# 3. Damping rings

The damping ring design presented in the TDR has already included the possibility of running at 10 Hz duty cycle to support the 10-Hz mode positron production scheme, albeit at the nominal beam current of 390 mA (1312 bunches). This requires reducing in the damping time by a factor of two as compared to the nominal 5 Hz operation, which is achieved by increasing the energy loss per turn by the same factor. The factor of two higher beam power requires additional RF power. For the luminosity upgrade, we consider two positron rings, both of which are therefore specified at the baseline parameters. However, the electron ring current doubles (780 mA), requiring a further factor of two RF beyond that discussed in the TDR. Table 2 summarises the relevant electron damping ring parameters.

Table 2: Key power parameters for the electron damping ring. (Grey shaded numbers were not considered in the TDR.)

| Rep rate | $n_b$ | $I_b$ (DR) mA | $\Delta E$ per turn GeV | $P_{beam}$ MW | # cavities | # klystorns | $P_{SR}$/m wiggler kW/m |
|---|---|---|---|---|---|---|---|
| 5 Hz | 1312 | 390 | 4.5 | 1.8 | 10 | 5 | 15.6 |
|  | 2625 | 780 | 4.5 | 3.6 | 12 | 6 | 31.2 |
| 10 Hz | 1312 | 390 | 6.1 | 2.4 | 10 | 5 | 21.1 |
|  | 2625 | 780 | 6.1 | 4.8 | 14 | 7 | 42.0 |

The current damping ring design allows space for up to 16 cavities. The high power loss in the 113 m wiggler sections will require special attention of the design of the vacuum systems and in particular the photon stops and dumps in those sections.

Finally, Table 3 gives an estimate of the total AC power required for the various operation modes of the damping rings (based on scaling of the baseline parameter numbers [3]). The TDR design currently facilitates up to ~22 MW of AC power, and 10 Hz operation at high beam current (2625 bunches) will require the installation of an additional ~8 MW of AC power.

Table 3: Estimate total AC power consumption for the Damping Rings.

| Rep rate | $n_b$ = 1312 2 rings | $n_b$ = 2625 3 rings |
|---|---|---|
| 5 Hz | 10.1 | 21.5 |
| 8 Hz | 13.1 | 25.6 |
| 10 Hz | 15.1 | 30.3 |

# 4. Positron source

The primary risk associated with increase beam power operation resides in the positron source, specifically with the power density on the rotating production target. There are three aspects which need to be considered with respect to the upgrade scenarios: (1) the increase in bunch number; (2) the increase in repetition rate; and (3) the exclusion of the 10-Hz mode positron production scheme at $E_{CM} \leq 300 \text{ GeV}$ when we consider the possibility of using 10 Hz for collisions.

Increasing the pulse repetition rate by up to a factor of two increases the average power deposition in the target, but not the peak energy deposition. This amounts to a possible higher average running temperature (still moderately low), additional water cooling, and over time a higher level of radiation damage, shortening the expected target lifetime. However these are not the primary issues and the effects are relatively benign. The primary constraint on the target parameters are to do with instantaneous energy deposition within a single bunch train, and the stresses that this can generate. The level of peak energy deposition density per pulse effectively scales as the beam current and so increases by a factor of 1.5 for the TDR luminosity upgrade. At 500 GeV centre of mass operation this results in approximately peak energy deposition density of ~100 J/g, resulting in an instantaneous peak temperature rise of some ~190 °C. One of the critical figures of merit is the peak mechanical stress that is induced; results of simple scaling suggest this would be 300-350 MPa, which remains below limits at which damage can occur, but with less than factor of two safety margin. Table 4 shows a summary of the target parameters for the various scenarios being considered. It should be noted that the mechanical and thermal estimates require further confirmation using more robust and detailed calculations using finite-element analysis methods[3]. In general the photon target R&D remains a high priority.

Table 4: Estimated (scaled) target parameters for the various operation modes considered. $\sigma_\gamma$ is the photon beam size on the target; ΔE peak is the estimate peak energy deposition per pulse; ΔT max is the estimate maximum instantaneous temperature rise, which remains relatively small across all the scenarios. The induced peak stress is assumed to scale with ΔT and remains below the reported fatigue strength of ~500 MPa (for $10^7$ cycles). The 250 GeV numbers are based on a ~200 m undulator [4].

| $E_{cm}$ | | 250 | | 350 | | 500 | |
|---|---|---|---|---|---|---|---|
| $E_{beam}$ | GeV | 125 | 125 | 175 | 175 | 250 | 250 |
| $n_b$ | | 1312 | 2625 | 1312 | 2625 | 1312 | 2625 |
| $\sigma_\gamma$ | mm | 2.5 | 2.5 | 1.2 | 1.2 | 0.8 | 0.8 |
| ΔE peak | J/g | 44 | 66 | 66 | 99 | 68 | 100 |
| ΔT max | K | 84 | 127 | 125 | 189 | 129 | 192 |
| stress (peak) | MPa | 140 | 211 | 209 | 315 | 215 | 319 |

At 250 GeV centre-of-mass, an alternative to the TDR 10-Hz mode positron production scheme needs to be found to allow 10 Hz collisions (point 3 above). This has been the subject of much recent study (see for example [4]), which have shown that a 120-125 GeV drive beam can be used to achieve the required positron yield, using a longer undulator (200-230 m as compared to the nominal TDR baseline of 147 m active length) as shown in Figure 3.

Space is already foreseen for the longer undulator in the TDR baseline design. The target parameters in Table 4 for 250 GeV are based on this option. Further possibilities include a shorter period undulator (reduced from the current prototype values of 11.5 mm) which is an on-going R&D topic.

---

[3] Several studies have already been made (see for example [4]) but much remains to be done.

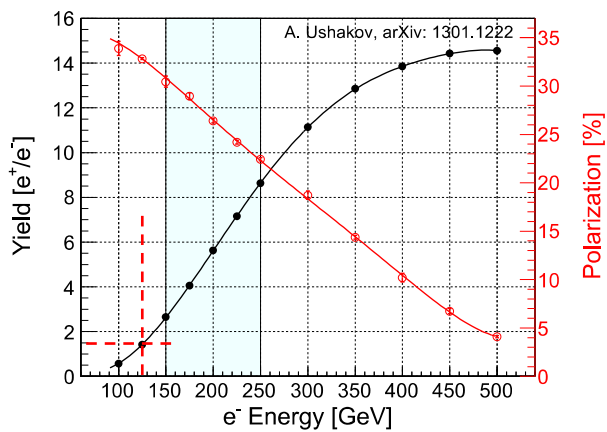

Figure 3: Yields achievable with a 231m undulator (K=0.92) [4]. With these parameters an e+/e- yield of 1.5 is achievable at a beam energy of 125 GeV. (In practice the undulator field strength would be reduced when the electron beam energy is above 125 GeV to maintain a fixed yield of 1.5).

# 5. Summary of luminosity performance and AC power

Table 5 summarises the various scenarios discussed above, in particular the expected luminosity and required AC power, assuming a fully constructed 500 GeV machine as proposed in the TDR.

Table 5: Summary of operation scenarios and upgrades. Further options indicate those scenarios which go beyond the parameters discussed in the TDR, but are still within the technical scope of the TDR.

|  |  | TDR reference | | | | Further options | |
|---|---|---|---|---|---|---|---|
|  |  | baseline | | | upgrade | | |
| Ecm | GeV | 250 | 350 | 500 | 500 | 250 | 350 |
| Rep. rate | Hz | 5 | 5 | 5 | 5 | 10 | 8 |
| Bunches/pulse |  | 1312 | 1312 | 1312 | 2625 | 2625 | 2625 |
| Total beam power | MW | 5.3 | 7.4 | 10.5 | 21.0 | 21.0 | 23.5 |
| AC power | MW | 122[a] | 121 | 163 | 204 | 187 | 204 |
| Luminosity | ×10$^{34}$ cm$^{-2}$s$^{-1}$ | 0.75 | 1.0 | 1.8 | 3.2 | 3.0 | 3.2 |

a) TDR baseline assumes 10-Hz mode for e+ production

For simplicity, only the high current options (2625 bunches per pulse) are shown for the further options. By effectively making use of the "spare power" when running the main linacs at reduced gradient (via increasing the repetition rate), allows a constant luminosity across the centre-of-mass energy range from the light Higgs (250 GeV) up to the maximum design energy (500 GeV). Despite the higher power required for the damping rings (Section 3), the total AC power does not exceed that of the 500 GeV machine (upgrade).

# 6. Light Higgs factory (staged approach)

The TDR outlines the possible construction of a 250 GeV centre-of-mass machine (approximate half the main linac length) as a possible first phase of a staged construction for the ILC. In this case, the gradient remains at 31.5 MV/m. Running with high beam current or higher repetition rate results in an increase in required AC power. The TDR quotes 129 MW for the first stage machine (with a luminosity of 0.75×10$^{34}$ cm$^{-2}$s$^{-1}$), which includes significant overhead for 10-Hz mode for e+ production. However, assuming that 10-Hz production mode can be avoided (Section 4), this power drops to ~100 MW. In principle we can achieve the same factor of 4 increase as described above, by assuming 10Hz collision rates and 2625 bunches per pulse. In this case the total AC power would be ~186 MW. However, 10 Hz operation at 31.5 MV/m requires some study, in particular for the cryogenic load.

# 7. Conclusion

High beam power scenarios have been investigated for 250 GeV and 350 GeV centre-of-mass operation, assuming the full construction of the 500 GeV machine. They have shown that by running the main linacs at lower gradients and increasing the repetition rates, constant luminosity can be established across the centre-of-mass energy range 250 GeV < E$_{CM}$ < 500 GeV. Assuming this option and the high beam current scenario outline in the TDR (luminosity upgrade) a factor of 4 in luminosity could be achieved at 250 GeV centre-of-mass (3×10$^{34}$ cm$^{-2}$s$^{-1}$). The maximum expected power consumption is ~200 MW.

For an initial phase light Higgs factor, the same luminosity can be achieved by with a large increase in AC power (186 MW versus ~100 MW).

The high power options discussed come with some increased risks, primarily in the sources (photon target) and damping rings (synchrotron radiation power). However these risk are believed to be manageable.